%% file: main.tex
\documentclass[conference]{IEEEtran}
\input{defs}
\IEEEoverridecommandlockouts
\usepackage{cite}
\usepackage{amsmath,amssymb,amsfonts}
\usepackage{algorithmic}
\usepackage{graphicx}
\usepackage{textcomp}
\usepackage{xcolor}
\def\BibTeX{{\rm B\kern-.05em{\sc i\kern-.025em b}\kern-.08em
    T\kern-.1667em\lower.7ex\hbox{E}\kern-.125emX}}

\usepackage{lipsum}
\begin{document}



\title{
\ourframework: Towards \underline{R}eLU-\underline{R}educed Neural \underline{Net}work for Two-party Computation Based Private Inference
\vspace{-2mm}
}

\author{

  \IEEEauthorblockN{
    Hongwu Peng\IEEEauthorrefmark{1},
    Shanglin Zhou\IEEEauthorrefmark{1},
    Yukui Luo\IEEEauthorrefmark{2},
    Nuo Xu \IEEEauthorrefmark{3}, 
    Shijin Duan \IEEEauthorrefmark{2}, 
    Ran Ran\IEEEauthorrefmark{3}, 
    Jiahui Zhao \IEEEauthorrefmark{1},\\
    Shaoyi Huang\IEEEauthorrefmark{1}, 
    Xi Xie\IEEEauthorrefmark{1},
    Chenghong Wang\IEEEauthorrefmark{4},
    Tong Geng\IEEEauthorrefmark{5},
    Wujie Wen\IEEEauthorrefmark{3},
    Xiaolin Xu\IEEEauthorrefmark{2},
    and Caiwen Ding\IEEEauthorrefmark{1}
  }

  \IEEEauthorblockA{\IEEEauthorrefmark{1}University of Connecticut 
  \IEEEauthorrefmark{2}Northeastern University
  \IEEEauthorrefmark{3}Lehigh University
  \IEEEauthorrefmark{4}Duke University
  \IEEEauthorrefmark{5}University of Rochester \\
    \IEEEauthorrefmark{1}\{hongwu.peng, shanglin.zhou, jiahui.zhao, shaoyi.huang, xi.xie, caiwen.ding\}@uconn.edu, \\
    \IEEEauthorrefmark{2}\{luo.yuk, duan.s, x.xu\}@northeastern.edu,
    \IEEEauthorrefmark{3}\{nux219, rar418, wuw219\}@lehigh.edu, \\
    \IEEEauthorrefmark{4}\{cw374\}@duke.edu,
    \IEEEauthorrefmark{5}tgeng@ur.rochester.edu
    \vspace{-3mm}
    }
}

\maketitle
\input{sections/0_abstract}


\input{sections/1_introduction}
\input{sections/2_background}
\input{sections/3_building_block}

\input{sections/4_accelerator}
\input{sections/4_main_problem_formulation}

\input{sections/4_main_archi_search}
\input{sections/6_evaluation}
\input{sections/7_conclusion}
\bibliographystyle{unsrt}
\vspace{-4pt}
\bibliography{ref}

\clearpage

\end{document}

%% file: defs.tex
\usepackage{xspace}
\usepackage{xcolor}
\usepackage{amsmath, amsfonts, amsthm} 
\usepackage{graphicx}
\usepackage{float} 
\usepackage{hyperref}
\usepackage{wrapfig}
\usepackage{mathrsfs}
\usepackage{multicol}
\usepackage{balance}
\usepackage{mathtools}
\usepackage{tikz}

\usepackage{tabularx} 
\usepackage{adjustbox} 
\usepackage{multirow} 

\usepackage{complexity}
\usepackage{relsize}
\usepackage{graphicx}
\usepackage{epsfig}
\usepackage{flushend}
\usepackage{epstopdf}
\usepackage{longtable}
\usepackage{stmaryrd}
\usepackage{pstricks}
\usepackage{pst-tree}
\usepackage{epstopdf}
\usepackage{lipsum}
\usepackage{color}
\usepackage{framed}
\usepackage[normalem]{ulem}
\usepackage{float}
\usepackage{caption}
\usepackage{fancyvrb}
\usepackage{bm}

\usepackage{algorithmic}
\usepackage{algorithm}

\newcommand{\eat}[1]{}

\newcommand{\ignore}[1]{}

\newcommand{\share}[1]{\llbracket #1 \rrbracket}

\newcommand{\pprotocol}[5]{
{
\begin{center}
 \advance\leftskip-2cm
 \advance\rightskip-2cm
\setlength{\protowidth}{\textwidth}
\addtolength{\protowidth}{\intextsep}
\fbox{
        \hbox{\quad
        \begin{minipage*}{\protowidth}
        #5
        \end{minipage*}
        \quad}
        }
        \caption{\label{#3} #2}
\end{center}

} }

\renewcommand{\paragraph}[1]{\vskip 0.08in\noindent {\bf #1.}}

\usepackage{subfig}

\newcommand{\ourframework}{RRNet\xspace}

%% file: sections/0_abstract.tex
\begin{abstract}
The proliferation of deep learning (DL) has led to the emergence of privacy and security concerns. To address these issues, secure Two-party computation (2PC) has been proposed as a means of enabling privacy-preserving DL computation. However, in practice, 2PC methods often incur high computation and communication overhead, which can impede their use in large-scale systems. To address this challenge, we introduce \ourframework, a systematic framework that aims to jointly reduce the overhead of MPC comparison protocols and accelerate computation through hardware acceleration. Our approach integrates the hardware latency of cryptographic building blocks into the DNN loss function, resulting in improved energy efficiency, accuracy, and security guarantees. Furthermore, we propose a cryptographic hardware scheduler and corresponding performance model for Field Programmable Gate Arrays (FPGAs) to further enhance the efficiency of our framework. Experiments show  \ourframework achieved a much higher ReLU reduction performance than all SOTA works on CIFAR-10 dataset. 
\end{abstract}

%% file: sections/1_introduction.tex
\section{\textbf{Introduction}}

Machine-Learning-as-a-Service (MLaaS) has emerged as a popular solution for accelerating inference in various applications~\cite{wu2021federated, wang2021lightweight, peng2022length, yang2020co, wu2021federated2, kan2022brain, wu2020intermittent, peng2022towards, huang2022automatic, huang2021hmc, qi2021accommodating}. 
The challenges of MLaaS comes from several folds: inference latency and privacy. To accelerate the MLaaS training and inference application, accelerated gradient sparsification~\cite{bao2022doubly, bao2022accelerated} and model compression methods~\cite{yuan2021improving, yan2022radars, huang2022dynamic, kan2022fbnetgen ,wu2020enabling, bao2019efficient, kan2021zero, luo2022codg, peng2021accelerating} are proposed. 
On the other side, a major limitation of MLaaS is the requirement for clients to reveal raw input data to the service provider, which may compromise the privacy of users. This issue has been highlighted in previous studies such as~\cite{kumar2020cryptflow}. In this work, we aim to address this challenge by proposing a novel approach for privacy-preserving MLaaS. Our method enables clients to maintain the confidentiality of their input data while still allowing for efficient and accurate inference.
Homomorphic Encryption (HE) is a powerful tool for securing small to medium-scale deep neural networks (DNNs) without incurring the high costs associated with bootstrapping or significant communication overhead. Other secure multiparty computation (MPC) protocols such as secret-sharing~\cite{knott2021crypten} and Yao's Garbled Circuits (GC)\cite{bellare2012adaptively} have also been proposed to support the evaluation of operator blocks in large-scale networks. However, our focus in this work is on the use of secure two-party computation (2PC)\cite{knott2021crypten} as a means of protecting DNN models.  


The main challenge in 2PC-based private inference (PI) is the overhead associated with the comparison protocol for non-linear operators~\cite{garay2007practical}. To address this challenge, existing works have focused on optimizing the cost of the ReLU operator by minimizing ReLU counts (e.g., DeepReduce~\cite{jha2021deepreduce}, CryptoNAS \cite{ghodsi2020cryptonas}) or replacing ReLUs with polynomials (e.g., CryptoNets~\cite{gilad2016cryptonets}, Delphi~\cite{mishra2020delphi}, SAFENet~\cite{lou2020safenet}). Another trend in the field has been the use of hardware acceleration for PI, such as using Graphics Processing Units (GPUs)~\cite{knott2021crypten, tan2021cryptgpu} to speed up MPC-based DNNs. However, both of these approaches have limitations in effectively exploring the design space of 2PC-based PI. In this work, we aim to address these limitations by proposing a novel approach for optimizing the cost of non-linear operators and hardware acceleration for PI that can effectively perform design exploration.

Current approaches for optimizing the performance of 2PC-based private inference (PI) rely on heuristic methods for evaluating the impact of different non-linear operators on system performance. In this work, we propose a novel approach, the \textbf{ReLU-Reduced Neural Architecture Search (\ourframework)} framework, that jointly optimizes the structure of the deep neural network (DNN) model and the hardware architecture to support high-performance MPC-based PI. Our framework eliminates the need for manual heuristic analysis by automating the process of exploring the design space and identifying the optimal configuration of DNN models and hardware architectures for 2PC-based PI.
We use FPGA accelerator design as a demonstration
and summarize our contributions:
\vspace{-1pt}




\leftmargini=4mm
\begin{enumerate}
\item  
We propose a novel approach to addressing the high computational cost of non-linear operators in 2PC-based PI. We introduce a trainable \textit{straight-through polynomial activation initialization} method that utilizes a trainable polynomial activation function as an alternative to the computationally expensive ReLU operator.
\item   
We develop a cryptographic hardware scheduler and performance model for FPGA platform. We also construct a latency lookup table to optimize scheduling of cryptographic operations for improved performance and energy efficiency.
\item   
We propose a differentiable  NAS framework that takes into account the constraints and latencies of cryptographic operators. Our framework enables the selection of appropriate polynomial or non-polynomial activation functions based on the specific needs of the task and the computational resources available. By integrating cryptographic considerations into the NAS process, our framework ensures that the resulting DNN models are both accurate and secure, while also being optimized for the target hardware platform.
\end{enumerate}
\vspace{-4pt}
 


%% file: sections/2_background.tex
\section{\textbf{Basic of Cryptographic Operators}}
\label{sec:Cry_Building_Block}

\subsection{\textbf{Secret Sharing}}

\noindent\textbf{2PC setup.} 
We consider a similar scheme involving two semi-honest in a MLaaS applications~\cite{demmler2015aby}, where two servers receive the confidential inputs from each other and invoke evaluation. 

%% file: sections/3_building_block.tex
\begin{figure}[t]
    \centering
      \includegraphics[width=1\linewidth]{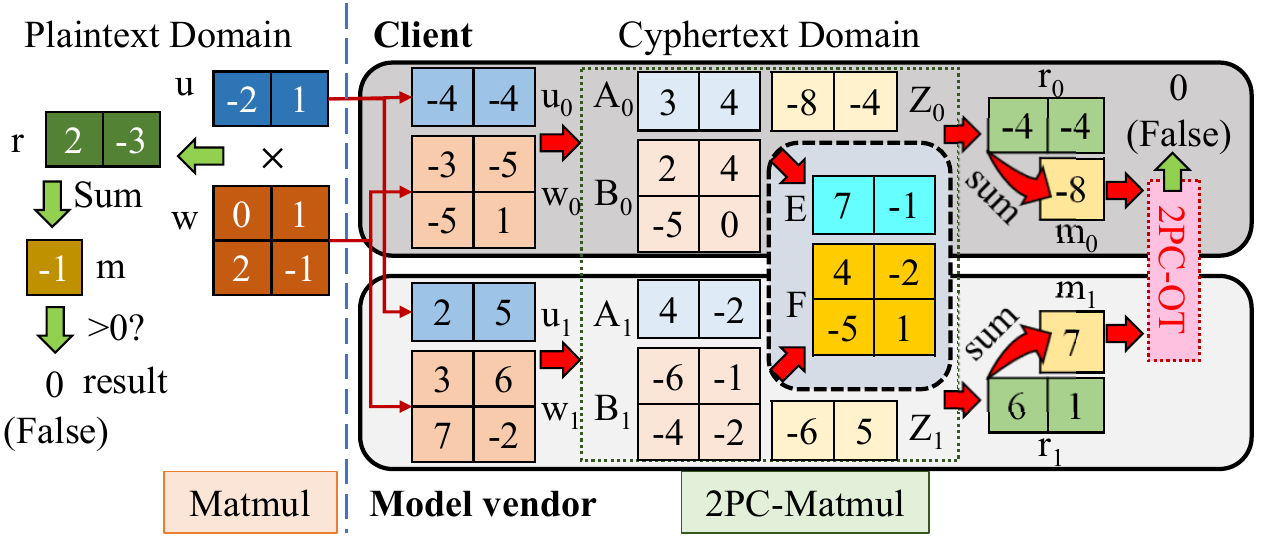}
      \vspace{-16pt}
  \captionof{figure}{A example of 4 bit plaintext vs. ciphertext evaluation. 
  }
  \vspace{-19pt}
    \label{fig:ss_example}
\end{figure}

\noindent\textbf{Additive Secret Sharing.} 
In this work, we evaluate 2PC secret sharing. 
As a symbolic representation, for a secret value $x\in \mathbb{Z}_m$, $\share{x}\gets(x_{S_0}, x_{S_1})$ denotes the two shares, where $x_{S_i}, i\in \{0,1\}$ belong to server $S_i$. Other notations are as below:
\begin{itemize}
    \item {\it Share Generation} $\mathbb{\textrm{shr}} (x)$: A random value $r$ in $\mathbb{Z}_{m}$ is sampled, and shares are generated as $\share{x}\gets (r, x-r)$.
    \item {\it Share Recovering}  $\mathbb{\textrm{rec}} ({\share{x}})$: Given shares $\share{x}\gets (x_{S_0}, x_{S_1})$, it computes $x\gets x_{S_0} + x_{S_1}$ to recover $x$.
\end{itemize}
An example of plaintext vs. secret shared based ciphertext evaluation is given in Fig.~\ref{fig:ss_example}, where ring size is 4 and $\mathbb{Z}_m =\{-8, -7, ... 7\}$. 
Details are given in following sections. 



\subsection{\textbf{Polynomial Operators Over Secret-Shared Data}}\label{sec:poly}
\noindent\textbf{Scaling and Addition.} We denote secret shared matrices as $\share{X}$ and $\share{Y}$. The encrypted evaluation is given in Eq.~\ref{eq:mat_ad_ss}. 
\vspace{-3pt}
\begin{equation}\label{eq:mat_ad_ss}
\share{aX+Y}\gets(aX_{S_0}+Y_{S_0}, aX_{S_1}+Y_{S_1})
\vspace{-6pt}
\end{equation}
\noindent\textbf{Multiplication.}
We consider the matrix multiplicative operations $\share{R}\gets \share{X} \otimes \share{Y}$ in the secret-sharing pattern.
We use oblivious transfer (OT)~\cite{kilian1988founding} based approach. 
To make the multiplicative computation secure, an extra Beaver triples~\cite{beaver1991efficient} should be generated as $\share{Z}=\share{A}\otimes\share{B}$, where $A$ and $B$ are randomly initialized. 
Specifically, their secret shares are denoted as $\share{Z}=(Z_{S_0}, Z_{S_1})$, $\share{A}=(A_{S_0}, A_{S_1})$, and $\share{B}=(B_{S_0}, B_{S_1})$. 
Later, two matrices are derived from given shares: $E_{S_i} = X_{S_i} - A_{S_i}$ and $F_{S_i} = Y_{S_i} - B_{S_i}$, in each party end separately. The intermediate shares are jointly recovered as $E\gets \mathbb{\textrm{rec}} {(\share{E})}$ and $F\gets \mathbb{\textrm{rec}} {(\share{F})}$. Finally, each party, i.e, server $S_i$, will calculate the secret-shared $R_{S_i}$ locally:
\vspace{-3pt}
\begin{equation}\label{eq:mat_mul_ss}
R_{S_i} = -i\cdot E \otimes F + X_{S_i} \otimes F  + E \otimes Y_{S_i} + Z_{S_i}
\vspace{-3pt}
\end{equation}



\noindent\textbf{Square.}
For the element-wise square operator shown $\share{R}\gets \share{X} \otimes \share{X}$, we need to generate a Beaver pair $\share{Z}$ and $\share{A}$ where $\share{Z}=\share{A}\otimes\share{A}$, and $\share{A}$ is randomly initialized. 
Then parties evaluate $\share{E}=\share{X} - \share{A}$ and jointly recover $E\gets \mathbb{\textrm{rec}} {(\share{E})}$. The result $R$ can be obtained through Eq.~\ref{eq:sq_eval}. 
\vspace{-3pt}
\begin{equation}\label{eq:sq_eval}
\vspace{-3pt}
R_{S_i} = Z_{S_i} + 2 E \otimes A_{S_i}  + E \otimes E
\end{equation}

\vspace{-10pt}
\subsection{\textbf{Non-Polynomial Operator Modules}}
\vspace{-3pt}
Non-polynomial operators such as ReLU and MaxPool are evaluated using secure comparison protocol. \\
\noindent\textbf{Secure 2PC Comparison.}
The 2PC comparison, a.k.a. millionaires protocol, is committed to determine whose value held by two parties is larger, without disclosing the exact value to each other. 
We adopt work~\cite{garay2007practical} for 2PC comparison. 

\begin{figure*}[ht]
    \centering
      \includegraphics[width=.98\linewidth]{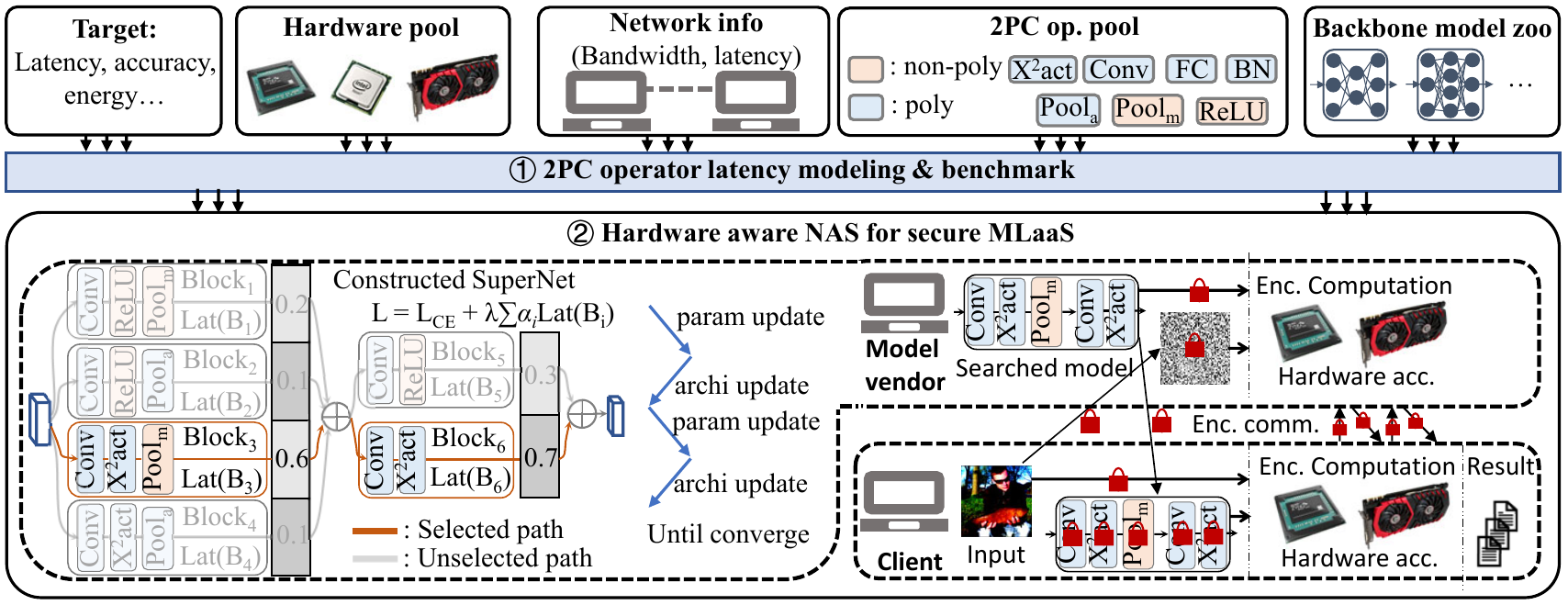}
  \captionof{figure}{Overview of \ourframework framework for 2PC DNN based private inference setup. 
  }
    \label{fig:NAS}
\end{figure*}

%% file: sections/4_accelerator.tex
\vspace{-2pt}
\section{\textbf{The \ourframework Framework}}
\vspace{-2pt}
The overview of the framework is given in Fig.~\ref{fig:NAS}. 
This section introduces the new cryptographic-friendly activation function and its initialization method. The modeling of DNN operators is conducted under 2PC setup on FPGA. As last, the hardware-aware NAS framework is proposed to find proper DNN architecture. 

%% file: sections/4_main_problem_formulation.tex
\vspace{-6pt}
\subsection{\textbf{Trainable $X^{2}act$ Non-linear Function.}}
\vspace{-2pt}
We use a hardware friendly trainable second order polynomial activation function as an non-linear function candidate, 
shown in Eq.~\ref{eq:x2act}, where $w_1$, $w_2$ and $b$ are all trainable parameters. 
We propose \textbf{\textit{straight through polynomial activation initialization} (STPAI)} method to set the $w_1$ and $b$ to be small enough and $w_2$ to be near to 1 in Eq.~\ref{eq:x2act} for initialization. 
\vspace{-5pt}
\begin{equation}\label{eq:x2act}
\delta(x) = \frac{c}{\sqrt[]{N_x}} w_1 x^2 + w_2 x + b
\vspace{-6pt}
\end{equation}

\subsection{\textbf{Search Space of Hardware-aware NAS.}}\label{sec:searchspace}
\vspace{-2pt}
We focus on convolutional neural networks (CNNs) in our study.
CNNs are mostly composed of Conv-Act-Pool and Conv-Act blocks. In work, we use the regular backbone model as a search baseline, such as the VGG family, mobilenetV3, and ResNet family. 
A toy example is shown in Fig.~\ref{fig:NAS}, where a two-layer supernet is constructed, and the first layer is Conv-Act-Pool, and the second layer is Conv-Act. The first layer has four combinations which are Conv-ReLU-Pool\textsubscript{m}, Conv-ReLU-Pool\textsubscript{a}, Conv-$X^{2}act$-Pool\textsubscript{m}, and Conv-$X^{2}act$-Pool\textsubscript{a}. The second layer has two combinations: Conv-ReLU and Conv-$X^{2}act$. The Conv block's parameters can be either shared among candidates or separately trained during the search. 
\vspace{-5pt}



%% file: sections/4_main_archi_search.tex
\subsection{\textbf{Differentiable Harware Aware NAS Algorithm}}
\label{sec:archi_search}

\begin{algorithm}[htb] 
\small
\caption{Differentiable Polynomial Architecture Search. 
} 
\label{alg:framework} 
\begin{algorithmic}[1] 
\REQUIRE 
$M_b$: backbone model; $D$: a specific dataset\\
~~~~ $Lat(OP)$: latency loop up table; $H$: hardware resource
\ENSURE 
Searched polynomial model $M_p$
\WHILE{not converged}
 \STATE Sample minibatch $x_{trn}$ and $x_{val}$ from trn. and val. dataset
 \STATE // Update architecture parameter $\alpha$:
 \STATE Forward path to compute $\zeta_{trn}(\omega, \alpha)$ based on $x_{trn}$
 \STATE Backward path to compute $\delta \omega = \frac{\partial \zeta_{trn}(\omega, \alpha)}{\partial \omega}$
 \STATE Virtual step to compute $ \omega' = \omega - \xi\delta\omega$
 \STATE Forward path to compute $\zeta_{val}(\omega', \alpha)$ based on $x_{val}$
 \STATE Backward path to compute $\delta \alpha' = \frac{\partial \zeta_{val}(\omega', \alpha)}{\partial \alpha}$
 \STATE Backward path to compute $\delta \omega' = \frac{\partial \zeta_{val}(\omega', \alpha)}{\partial \omega'}$
 \STATE Virtual steps to compute $ \omega^{\pm} = \omega \pm \varepsilon\delta\omega'$ 
 \STATE Two forward path to compute $\zeta_{trn}(\omega^{\pm}, \alpha)$ 
 \STATE Two backward path to compute $\delta \alpha^{\pm} =  \frac{\partial \zeta_{trn}(\omega^{\pm}, \alpha)}{\partial \alpha}$ 
 \STATE Compute hessian $\delta \alpha'' = \frac{\delta \alpha^{+} - \delta \alpha^{-}}{2\varepsilon}$
 \STATE Compute final architecture parameter gradient $\delta\alpha = \delta \alpha'- \xi\delta \alpha'' $
 \STATE Update architecture parameter using $\delta\alpha$ with Adam optimizer
 \STATE // Update weight parameter $\omega$:
 \STATE Forward path to compute $\zeta_{trn}(\omega, \alpha)$ based on $x_{trn}$
 \STATE Backward path to compute $\delta \omega = \frac{\partial \zeta_{trn}(\omega, \alpha)}{\partial \omega}$
 \STATE Update architecture parameter using $\delta\omega$ with SGD optimizer
\ENDWHILE
\\ Obtain architecture by $OP_{l}(x) = OP_{l, k^*}(x), \: s.t. \: k^* = \mathbb{\textrm{argmax}}_{k} \: \theta_{l, k}$
\end{algorithmic}
\end{algorithm}

In this work, we incorporate latency constraint into the target loss function of the DARTS framework~\cite{liu2018darts}, and develop a differentiable cryptographic hardware-aware micro-architecture search framework. We firstly determine a supernet model for NAS, and introduces gated operators $OP_{l}(x)$ which parametrizes the candidate operators $OP_{l, j}(x)$ selection with a trainable weight $\alpha_{l,k}$ (Eq.~\ref{eq:softmax_para}). For example, a gated pooling operator consists of MaxPool and AvgPool operators and 2 trainable parameters for pooling selection. The latency of the operators could be determined based on performance predictor. A parameterized latency constraint is given as $Lat(\alpha) = \sum_{l = 1}^{n}\sum_{j = 1}^{m}\theta_{l, j}Lat(OP_{l, j})$, 
where the latency of gated operators are weighted by $\theta_{l, j}$. 
We incorporate the latency constraint into the loss function as $\zeta(\omega, \alpha) = \zeta_{CE}(\omega, \alpha) + \lambda Lat(\alpha)$, and penalize the latency $Lat(\alpha)$ by $\lambda$. 




\vspace{-10pt}
\begin{equation}\label{eq:softmax_para}
\theta_{l, j} = \frac{\exp(\alpha_{l,j})}{\sum_{k = 1}^{m}\exp(\alpha_{l,k})}, \: OP_{l}(x) = \sum_{k = 1}^{m} \theta_{l, k}OP_{l, k}(x)
\vspace{-6pt}
\end{equation}

\begin{figure*}[h!]
    \centering
    \centering
\begin{multicols}{2}
\subfloat [\label{fig:hardware_throughput}Searched model accuracy comparison]   {\includegraphics[width=0.92\columnwidth]{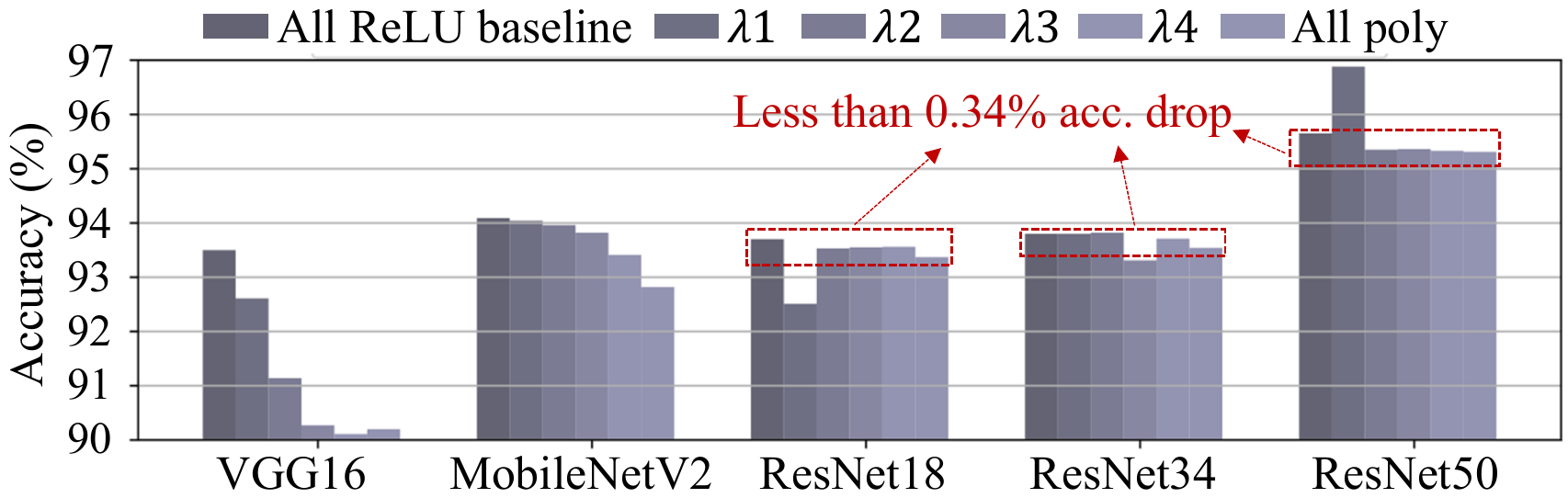}\par }
\subfloat  [\label{fig:attention_throughput}Searched model private inference latency comparison]  {\hspace{.3in}\includegraphics[width=0.9\columnwidth]{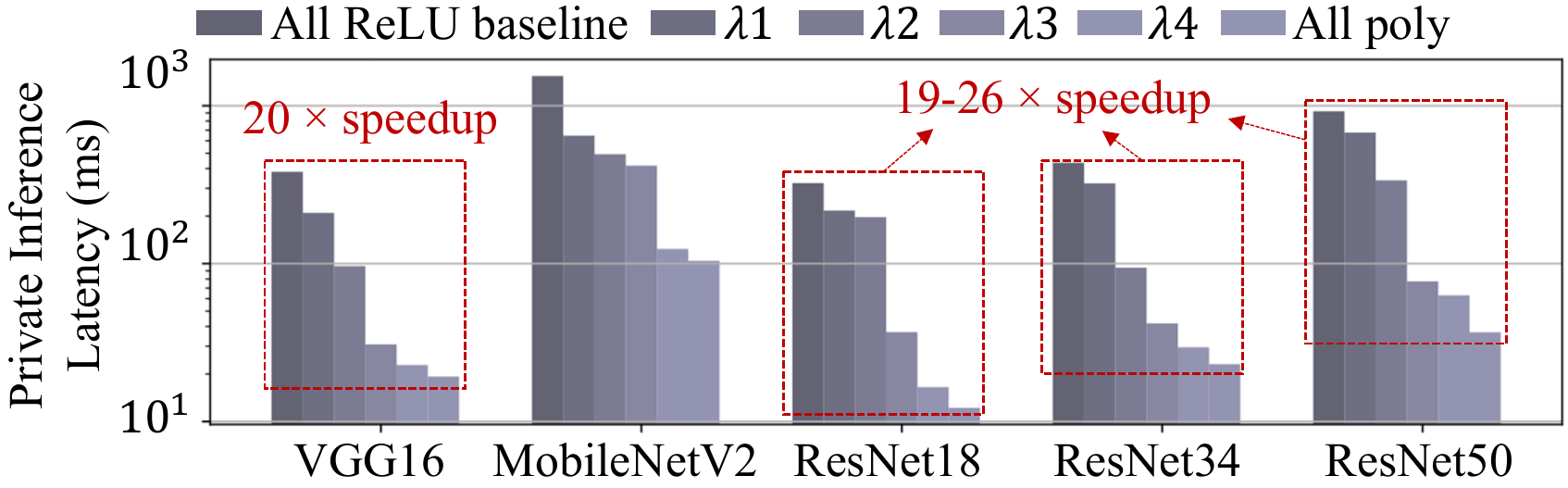}\par }
\end{multicols}
    \caption{\ourframework framework evaluation under 2PC PI setup. Network banwidth: 1 GB/s. Device: ZCU104. }
    \label{fig:pasnet_cifar10}
\end{figure*}

The optimization objective of our design is shown in Eq.~\ref{eq:optimization}, we aim to minimize the validation loss $\zeta_{val}(\omega^*, \alpha)$ with regard to architecture parameter $\alpha$. The optimal weight $\omega^*$ is obtained through minimize the training loss. The second order approximation of the optimal weight is given as $\omega^* \approx \omega' = \omega - \xi \: \delta \zeta_{trn}(\omega, \alpha)/\delta\omega$,
the approximation is based on current weight parameter and its' gradient. The virtual learning rate $\xi$ can be set equal to that of weight optimizer. 


\vspace{-12pt}
\begin{equation}\label{eq:optimization}
\mathbb{\textrm{argmin}}_{\alpha} \: \zeta_{val}(\omega^*, \alpha), \: s.t. \: \omega^* = \mathbb{\textrm{argmin}}_{\omega} \: \zeta_{trn}(\omega, \alpha)
\vspace{-4pt}
\end{equation}
Eq.~\ref{eq:opt_2} gives the approximate $\alpha$ gradient using chain rule, the second term of $\alpha$ gradient can be further approximated using small turbulence $\varepsilon$, where weights are $\omega^{\pm} = \omega \pm \varepsilon \: \delta\zeta_{val}(\omega', \alpha)/\delta\omega'$ 
and Eq.~\ref{eq:hession} is used for final $\alpha$ gradient. 
\vspace{-4pt}
\begin{equation}\label{eq:opt_2}
\delta\zeta_{val}(\omega', \alpha)/\delta\alpha - \xi  \: \delta\zeta_{val}(\omega', \alpha)/\delta\omega' \: \delta\delta\zeta_{trn}(\omega, \alpha)/\delta\omega\delta\alpha
\vspace{-4pt}
\end{equation}
\begin{equation}\label{eq:hession}
\frac{\delta\delta\zeta_{trn}(\omega, \alpha)}{\delta\omega\delta\alpha}
 =  \delta(\zeta_{trn}(\omega^+, \alpha) - \zeta_{trn}(\omega^-, \alpha))/(2\varepsilon\delta\alpha)
 \vspace{-2pt}
\end{equation}

With the help of analytical modeling of optimization objective, we are able to derive the differentiable polynomial architecture search framework in Algo.~\ref{alg:framework}. The input of search framework includes backbone model $M_b$, dataset $D$, latency loop up table $Lat(OP)$, and hardware resource $H$. The algorithm returns a searched polynomial model $M_p$. The algorithm iteratively trains the architecture parameter $\alpha$ and weight $\omega$ parameter till the convergence. Each $\alpha$ update requires 4 forward paths and 5 backward paths according to Eq.~\ref{eq:optimization} to Eq.~\ref{eq:hession}, and each $\omega$ update needs 1 forward paths and 1 backward paths. After the convergence of training loop, the algorithm returns a deterministic model architecture by applying $OP_{l}(x) = OP_{l, k^*}(x), \: s.t. \: k^* = \mathbb{\textrm{argmax}}_{k} \: \alpha_{l, k}$. The returned architecture is then used for 2PC based PI evaluation. 

%% file: sections/6_evaluation.tex
\section{\textbf{Evaluation}}
\noindent\textbf{Hardware setup.} Our experiment platform is based on two ZCU104 MPSoCs, both are connected to a router with $Rt_{bw} = 1 GB/s$ through LAN. The load/store bus width is 128-bit and our data is 32-bit, thus, we simultaneously load and store four data and implement the kernel on $freq = 200MHz$. Fixed point ring size is set as 32 bits for 2PC inference.

\noindent\textbf{Datasets and Backbone Models. }
We evaluate \ourframework on two public datasets: CIFAR-10~\cite{krizhevsky2009learning} and ImageNet~\cite{krizhevsky2012imagenet} for image classification tasks.



\noindent\textbf{Systems Setup. }
All polynomial architecture search experiments are conducted in plaintext domain on Ubuntu 18.04 and Nvidia Quadro RTX 6000 GPU with 24 GB GPU memory. 
The cryptographic DNN inference experiment is conducted on an FPGA-based accelerator for 2PC DNN setup. Two ZCU104 boards are used for server 0 and server 1, which are equipped with XCZU7EV MPSoC for the PS-PL system. Two boards are connected to a router with the Ethernet LAN setup. The FPGA accelerators are optimized with coarse-grained and fine-grained pipeline structures. 

\vspace{-2pt}
\subsection{\textbf{Hardware-aware NAS Evaluation}}
Our hardware-aware NAS experiment (algorithm descripted in Sec.~\ref{sec:archi_search}) was conducted on CIFAR-10 training dataset. A new training \& validation dataset is randomly sampled from the CIFAR-10 training dataset with 50\%-50\% split ratio. 

The hardware latency is modeled through FPGA performance predictor, and the $\lambda$ for latency constraint in loss function is tuned to generate architectures with different latency-accuracy trade-off. 
Prior search starts, the major model parameters are randomly initialized, and the polynomial activation function is initialized through \textbf{STPAI} method. 
We use VGG-16~\cite{simonyan2014very}, ResNet-18, ResNet-34, ResNet-50~\cite{he2016deep}, and MobileNetV2~\cite{sandler2018mobilenetv2} as backbone model structure to evaluate our \ourframework framework. 

The finetuned model accuracy under 2PC setting with regard to $\lambda$ setting can be found in Fig.~\ref{fig:pasnet_cifar10}(a). The baseline model with all ReLU setting and all-polynomial operation based model are also included in the figure for comparison. Generally, a higher polynomial replacement ratio leads to a lower accuracy. The VGG-16 model is the most vulnerable model in the study, while the complete polynomial replacement leads to a 3.2\% accuracy degradation (baseline 93.5\%). On the other side, ResNet family are very robust to full polynomial replacement and there are only $0.26\%$ to $0.34\%$ accuracy drop for ResNet-18 (baseline 93.7\%), ResNet-34 (baseline 93.8\%) and ResNet-50 (baseline 95.6\%). MobileNetV2's is in between the performance of VGG and ResNet, in which a full polynomial replacement leads to $1.27\%$ degradation (baseline 94.09\%).

On the other hand, Fig.~\ref{fig:pasnet_cifar10}(b) presents the latency profiling result of searched models performance on CIFAR-10 dataset under 2PC setting. All polynomial replacement leads to 20 times speedup on VGG-16 (baseline 382 ms), 15 times speedup on MobileNetV2 (baseline 1543 ms), 26 times speedup, ResNet-18 (baseline 324 ms), 19 times speedup on ResNet-34 (baseline 435 ms), and 25 times on speedup ResNet-50 (baseline 922 ms). With most strict constraint $\lambda$, the searched model latency is lower.  

\vspace{-2pt}
\subsection{\textbf{Cross-work ReLU Reduction Performance Comparison}}
\vspace{-2pt}

\begin{figure}[htpb!]
\vspace{-5pt}
\centering
\includegraphics[width=0.9\linewidth]{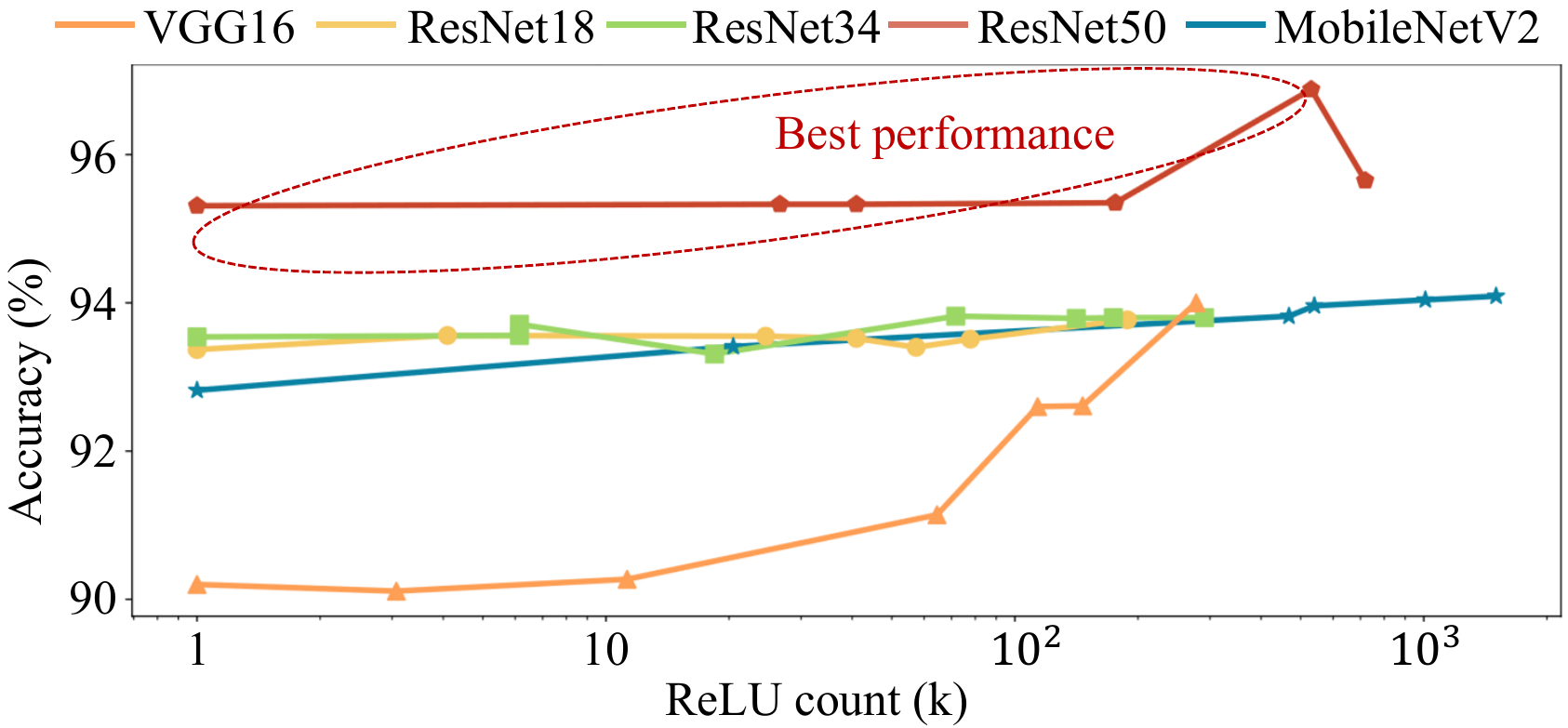}
\vspace{-3pt}
\caption{Accuracy-ReLU count trade-off on CIFAR-10. }
\label{fig:nas_acc_pareto}
\vspace{-12pt}
\end{figure}

\begin{figure} [htpb!]
\centering
\includegraphics[width = 0.9\linewidth]{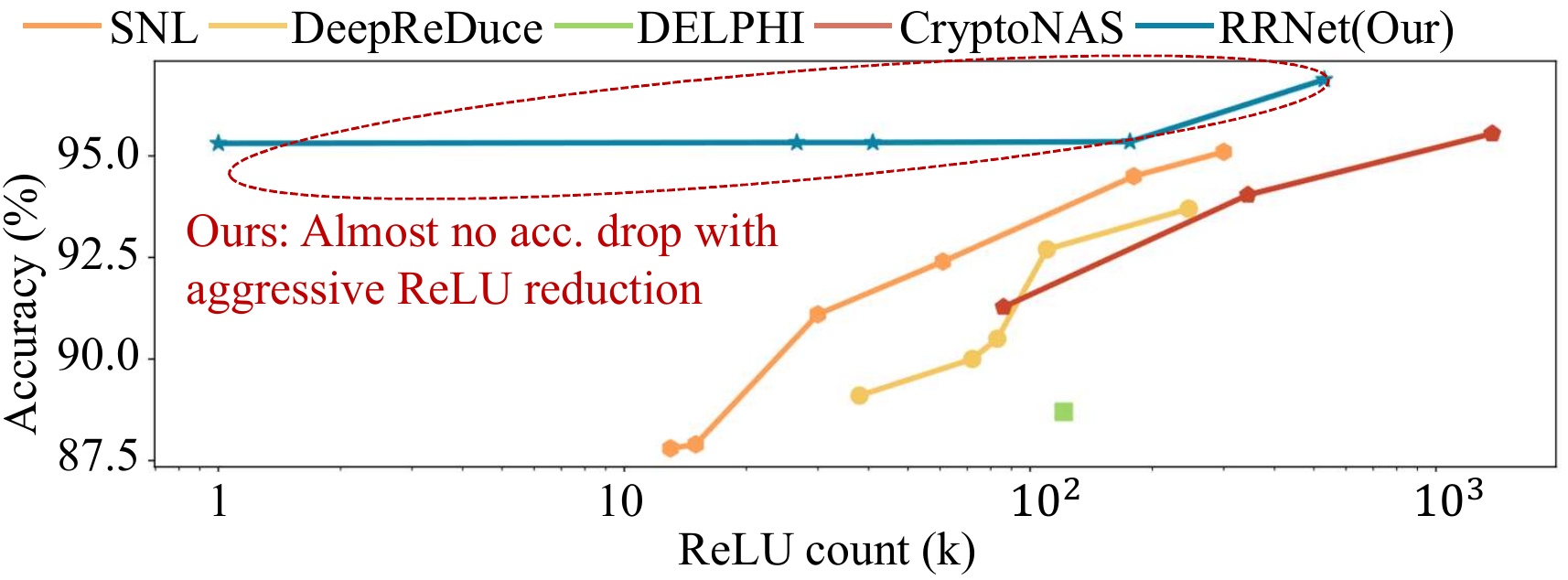}
\caption{ReLU reduction comparison on CIFAR-10. }
\label{fig:algo_cross_work}
\vspace{-7pt}
\end{figure}

A futher accuracy-ReLU count analysis is conducted and compared with SOTA works with ReLU reduction: DeepReDuce \cite{jha2021deepreduce}, DELPHI \cite{mishra2020delphi}, CryptoNAS \cite{ghodsi2020cryptonas}, and SNI \cite{cho2022selective}. As shown in Fig.~\ref{fig:nas_acc_pareto}, we generate the pareto frontier with best accuracy-ReLU count trade-off from our architecture search result. We name the selected models as \textbf{\ourframework}, and compare it with other works. The accuracy-ReLU count comparison is show in Fig.~\ref{fig:algo_cross_work}. Our work achieves a much better accuracy vs. ReLU comparison than existing works, especially at the situation with extremely few ReLU counts. 

%% file: sections/7_conclusion.tex
\vspace{-4pt}
\section{\textbf{Conclusion}}
\vspace{-4pt}
In the work, to reduce the high comparison protocol overhead from the non-linear operators in 2PC-based privacy-preserving DL implementation, we propose the \ourframework framework that enables low latency, high energy efficiency \& accuracy 2PC-DL. 
Experiments show  \ourframework achieved a much higher ReLU reduction performance than all SOTA works on CIFAR-10 dataset. 